\documentclass[sigconf,authorversion,nonacm]{acmart}

\AtBeginDocument{%
  }

\setcopyright{acmcopyright}
\copyrightyear{2022}
\acmYear{2022}
\acmDOI{XXXXXXX.XXXXXXX}

\acmConference[ICAIF-22]{The 3rd ACM International Conference on AI in Finance}{November 02--04, 2022}{New York City}
\acmPrice{15.00}
\acmISBN{978-1-4503-XXXX-X/18/06}

\usepackage{multicol}
\usepackage{multirow}
\usepackage{subfigure}

\begin{document}

\title[BERT-based Financial Sentiment]{BERT-based Financial Sentiment Index and LSTM-based Stock Return Predictability}

\author{Joshua Zoen-Git Hiew}
\authornote{Both authors contributed equally to this research.}
\email{joshuazo@ualberta.ca}
\affiliation{%
  \institution{University of Alberta}
  \city{Edmonton}
  \country{Canada}
}
\author{Xin Huang}
\authornotemark[1]
\email{huangxin@se.cuhk.edu.hk}
\affiliation{%
  \institution{The Chinese University of Hong Kong}
  \city{Hong Kong}
  \country{P.R.C.}
}

\author{Hao Mou}
\email{mouhao@datastory.com.cn}
\affiliation{%
  \institution{DataStory}
  \city{Guangzhou}
  \country{P.R.C.}
}

\author{Duan Li}
\authornote{Deceased.}
\email{dli226@cityu.edu.hk}
\affiliation{%
  \institution{City University of Hong Kong}
  \city{Hong Kong}
  \country{P.R.C.}
}

\author{Qi Wu}
\email{qiwu55@cityu.edu.hk}
\affiliation{%
  \institution{City University of Hong Kong}
  \city{Hong Kong}
  \country{P.R.C.}
}

\author{Yabo Xu}
\email{arber@datastory.com.cn}
\affiliation{%
  \institution{DataStory}
  \city{Guangzhou}
  \country{P.R.C.}
}

\renewcommand{\shortauthors}{Hiew et al.}

\begin{abstract}
  Traditional sentiment construction in finance relies heavily on the dictionary-based approach, with a few exceptions using simple machine learning techniques such as Naive Bayes classifier. While the current literature has not yet invoked the rapid advancement in the natural language processing, we construct in this research a textual-based sentiment index using a well-known pre-trained model BERT developed by Google, especially for three actively trading individual stocks in Hong Kong market with at the same time the hot discussion on Weibo.com. On the one hand, we demonstrate a significant enhancement of applying BERT in financial sentiment analysis when compared with the existing models. On the other hand, by combining with the other two commonly-used methods when it comes to building the sentiment index in the financial literature, i.e., the option-implied and the market-implied approaches, we propose a more general and comprehensive framework for the financial sentiment analysis, and further provide convincing outcomes for the predictability of individual stock return by combining LSTM (with a feature of a nonlinear mapping). It is significantly distinct with the dominating econometric methods in sentiment influence analysis which are all of a nature of linear regression.
\end{abstract}

\begin{teaserfigure}
  \includegraphics[width=\textwidth]{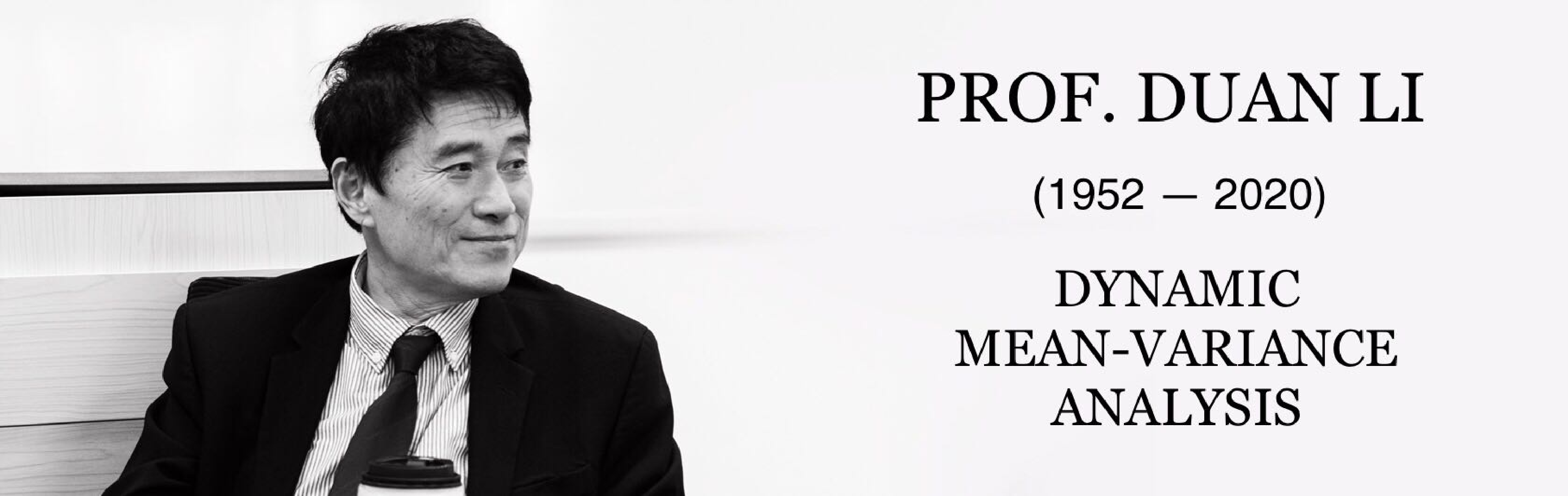}
  \caption{In memory of Prof. Li (\url{https://www.se.cuhk.edu.hk/obituary-for-prof-li-duan/}).}
  \label{fig:teaser}
\end{teaserfigure}

\maketitle

\section{Introduction}

It is a common belief that investors' sentiment is one of the important driving sources behind the financial market movement. Although the classical financial theory hypothesizes that investors are rational, extensive studies have already revealed the significant influence of their \emph{irrational} behavior, like optimistic or pessimistic sentiment (see \citet{LeeShleiferThaler} and \citet{BakerWurgler} among others). However, different research works adopt different sentiment measures. Sending questionnaires to investors is a very classical way to collect public opinion upon the market environment and market trend. The obvious drawback of this approach, however, is a low frequency on data acquisition, since a survey is usually conducted once a week, a month, or even a quarter. For instance, the sentiment proxy in \citet{BrownCliff} is based on weekly survey data from the American Association of Individual Investors. Some quantitative methods are then proposed. For example, both \citet{BakerWurgler} and \citet{ChongCaoWong} apply the principal component analysis (PCA) on a series of data set of selected market factors to extract the market-implied sentiment index, while \citet{Han} measures the investor sentiment from option-implied information. These methods focus on finding a proxy of sentiment leading to an indirect measure, compared with those approaches that directly deal with sentimental texts from the internet, like Twitter \citep[see][]{bollen2011twitter}, news or analysts' articles \citep[see][]{chen2018textual}. A couple of years ago, \citet{kearney2014textual} provided a comprehensive survey that summarizes different information sources, content analysis methods, and empirical models for the textual sentiment. As a conclusion, however, they suggest to extend the lexicons for textual content analysis, ignoring the rapid development in the natural language process (NLP). Although the sentiment analysis is a common research field in both machine learning (ML) and behavioral finance, there is still a big gap to integrate the research strength of these two.

In this paper, we construct a textual-based sentiment index by adopting the newly-devised NLP tool BERT from \citet{devlin2018bert} to posts that are published on the Chinese social media called Weibo, which represents the first attempt in the literature to apply this state-of-the-art learning model to the financial sentiment extraction. At this stage, our analysis focuses mainly on the \emph{individual} stock level. Namely, we mainly investigate three actively-trading listed companies in Hong Kong Stock Exchange (HKSE) in this pilot study, and they are Tencent (0700.HK), CCB (0939.HK), and Ping An (2318.HK). The reason why we select these three is that all of them possess sufficient exposures on Weibo so that we have a plenty of textual sources. We then demonstrate a better performance of BERT on sentiment construction compared with the other well-known models such as Multichannel Convolutional Neural Network (CNN) of \citet{kim2014convolutional} and Transformer of \citet{vaswani2017attention}, among others. Besides, through combining the BERT-based sentiment index with other two types of sentiment indices from the option-implied information and PCA on market data for the above three stocks, we next provide a deeper and more general financial sentiment analysis. More specifically, our BERT-based sentiment reflects more about individual investors' opinion, whereas the option-implied one followed by \citet{Han} represents more about the institutions' attitude. We would like to see how these two counterparts influence the market, together with the attendance of market-based index which is treated as an overall market sentiment. Finally, we address the stock return predictability by integrating sentiment indices that are constructed from different information sources. This is done by applying the powerful sequential neural network model Long Short-Term Memory (LSTM), in constrast to the classical econometrics tool like Vector Autoregression (VAR). Note that, as \citet{yan2018parsimonious} pointed out, LSTM model on quantile regression outperforms those traditional time-series analysis tools that are commonly used in the financial literature.

The rest of our paper is organized as follows. We first construct the textual sentiment index by BERT and also by other NLP models for comparison in Section 2. We then carry out a general financial sentiment analysis based on three different information sources in Section 3. We address the stock return predictability issue at the individual level in Section 4. Finally we conclude our paper in Section 5.

\section{Textual sentiment index construction}

In this paper, we focus on the individual-level sentiment analysis and take three listed firms in HKSE to conduct our experiment. More precisely, we select Tencent (0700.HK), Ping An (2318.HK), and CCB (0939.HK) as our individual stocks, and grab posts related to these three companies, respectively, during the time period from January 1, 2016 to December 31, 2018 on a daily basis from Weibo, a popular instant social media in China. In the following, we introduce our procedure of sentiment index construction and make evaluations among different ML models.

\subsection{Pre-processing work}\label{2.1}

After grabbing \emph{all} the firm-specific data from Weibo during the time period stated above, we start with a pre-processing work, which consists of cleaning and labelling on all posts. To filter out noisy posts like advertisements or others that are published by \emph{water army}\footnote{Internet water army, always sponsored by certain business entities, is a group of paid posters who post biased content for particular purposes and have flooded the social networks nowadays, as pointed out by \citet{chen2013battling}.}, we also adopt a detection model\footnote{A commercial software, launched by DataStory (www.datastory.com.cn), that is used to detect water army and has successfully served for more than 100 internet companies.} through labelling those jam information, at the same time when we label real sentimental posts about corresponding stocks. Note that in traditional sentiment analysis under supervised-learning framework, one may label a piece of context by emotional words, like ``happy'' and ``anger''. However, when it comes to financial texts, we prefer using the \emph{polarity}, i.e., ``positive'', ``negative'', and ``neutral'' to label the data, as they represent the attitude of posters corresponding to ``bullish'', ``bearish'', and ``ambiguous'' markets, respectively. Furthermore, we adopt a \emph{voting strategy}, similar as in \citet{ribeiro2016sentibench}, to enhance our accuracy of labelling. That is, we totally conduct six rounds of labelling and in each round, each post is manually labelled by at least three experts and we only keep those answers that are agreed by at least two of them. Moreover, we also need to achieve consensus on those conflicted posts among us before the next round starts. Eventually, we end up with 117,029 original posts for three stocks in total from Weibo during the considered time period mentioned above and randomly label 10,165 of them (8.69\%) that are ready to train and assess different ML models in the following.

\subsection{Evaluation by BERT and comparison with other ML models}

Proposed by \citet{devlin2018bert}, BERT, as an open-source model\footnote{See https://github.com/google-research/bert.}, is \emph{pre-trained} with massive datasets to encode bi-directional contexts through multi-layer transformers, and has been reported to achieve the state-of-the-art results in NLP downstream tasks. For instance, it completes the Stanford Sentiment Treebank (SST-2) task, as one of the General Language Understanding Evaluation (GLUE) benchmarks, with accuracy as high as 94.9\%. In this paper, we rely on the Chinese version ``BERT-Base, Chinese'' to do the fine-tuning.

To tackle the sentiment analysis as a basic \emph{text classification} task, there actually exist other genres of ML models. Formerly, researchers tend to use support-vector networks (SVM) \citep[see][]{Cortes95support-vectornetworks} or ensemble methods \citep[see][]{opitz1999popular} to build a classification model, while simpler lexical-based approaches are always adopted in finance, as surveyed in \citet{kearney2014textual}. With much more embedding methods and significant increase on computing power nowadays, deep learning methods are growingly dominating those statistical learning ones in all aspects of NLP. As a comparison with BERT, we mainly consider the other four famous models, i.e., the Recurrent Neural Network (RNN) based Bidirectional Long Short-Term Memory (BiLSTM) \citep[see][]{hochreiter1997long}, the Multichannel Convolutional Neural Network (CNN) \citep[see][]{kim2014convolutional}, the CPU-efficient FastText \citep[see][]{joulin2016bag} that is adopted by Facebook, and the Transformer with attention mechanism \citep[see][]{vaswani2017attention}. Note that since BiLSTM and Multichannel CNN are suggested to initialize with pre-trained word embedding like Shifted Positive pointwise mutual information (PMI) proposed by \citet{levy2014neural}, we finally take the PMI-enhanced versions of them \citep[see][for example]{li2018analogical}.

\begin{table*}[!htbp]
\caption{Comparison of performance across different models}
\centering
\begin{tabular}{lccc}
\toprule
ML Model                                & Precision\_micro & Recall\_micro & F1\_micro\\
\midrule
\textbf{BERT}                           & \textbf{79.3}    & \textbf{75.4} & \textbf{78.5}\\
Transformer + attention                 & 77.6             & 64.8          & 71.3          \\
PMI + Multichannel CNN                  & 75.9             & 60.6          & 64.3          \\
PMI + BiLSTM                            & 75.3             & 56.2          & 62.6          \\
FastText                                & 72.1             & 48.7          & 61.5          \\
\bottomrule
\end{tabular}
\label{model_evaluation}
\end{table*}

We split our labelled dataset described in Subsection \ref{2.1} into a training set and test set by a ratio of 80\% and 20\%, and a 10-fold cross validation is performed on the training set for all models. Table \ref{model_evaluation} shows performance evaluation for all selected models that are trained by the same labelled Weibo-post dataset mentioned in Subsection \ref{2.1}. In order to avoid impact from imbalanced proportion for different categories of our labelled result (15\% positive, 78\% neutral, and 7\% negative), we use \emph{micro-average} method to calculate the precision, the recall, and an averaged F1 score, respectively, as the common criteria for model evaluation. From the table we can see that BERT is superior across all indicators in our training process, especially on its significantly dominating recall rate even with the presence of its better precision. The above outcome demonstrates its strong capability over the other ML models for financial sentimental texts classification in Chinese, leading to the \emph{first} BERT-based financial sentiment index in the literature presented in the next subsection.

\subsection{BERT-based sentiment index}

We apply our fine-tuned BERT model to all unlabelled posts filtered by our detection model and classify them into three categories of polarity. Note that we treat those posts that are published after the trading time (4 p.m. (GMT+8) for Hong Kong market) as the influence for the next trading day, and calculate a BERT-based sentiment value $BSI_t^i$ for stock $i$ on a trading-day basis through
\begin{equation}
	BSI_t^i = \frac{Pos_t^i - Neg_t^i}{Pos_t^i + Neu_t^i + Neg_t^i}
\end{equation}
where $Pos_t^i$, $Neu_t^i$ and $Neg_t^i$ are the number of positive, neutral and negative texts that are related to stock $i$ and outputted by BERT on the trading day $t$, respectively.  Then, all $BSI_t^i$'s time-series data form our BERT-based financial sentiment index $BSI^i$ for stock $i$.

\section{Financial sentiment analysis based on different information channels}

Apart from the above textual channel that extracts financial sentiment from the social network by NLP techniques, there exist another two types of information sources that have been commonly utilized in the finance community. One of them is the \emph{risk-neutral implied skewness} which is based on the option price \citep[for example,][]{Han}, and this leads to the \emph{option-implied} sentiment; and another channel is through the \emph{market data} \citep[for example,][]{BakerWurgler}, resulting in the \emph{market-implied} sentiment. In our investigation, we construct these additional two sentiment indices and then take into consideration all these three indices in hand to conduct a more general financial sentiment analysis as follows.

\subsection{Option-implied and market-implied financial sentiment}

\citet{Han} discovers the relationship between option volatility smile, risk-neutral skewness and market sentiment. He finds that when the market tends to be bearish (or bullish), the slope of option volatility smile becomes steeper (or flatter) and the risk-neutral skewness changes to be more negative (positive). Accordingly, \citet{Han} proposes an option-implied sentiment proxy. Following his work, we construct the same sentiment index for our selected individual stocks using the implied skewness of their option information, respectively. And this is denoted by $OSI^i$.

Market data offer another traditional source for extracting market-oriented sentiment. \citet{BakerWurgler} identify a set of market data which they believe is driven by the investors' sentiment, and form an underlying proxy for such data set. They apply the \emph{principal component analysis} (PCA), which is sometimes considered as an unsupervised machine learning method, to extract this market-type sentiment. However, due to the low frequency of part of their selected market characteristics, such as the number of initial public offerings (IPO) within a month, \citet{ChongCaoWong} consider another set of market data which could represent the investors' sentiment on a daily basis. Therefore, in order to be in line with our previous sentiment indices, we choose to follow the work in \citet{ChongCaoWong} which concentrates on Hong Kong market as well and construct our market-implied sentiment index, denoted by $MSI^i$, for each individual stock.

The methodology of calculating the risk-neutral skewness, which is used to construct the option-implied sentiment index, can be found in \citet{BakshiKapadiaMadan}. The selected market characteristics that is adopted to build the market-implied sentiment index are illustrated in Appendix.

\subsection{Framework of financial sentiment analysis from three channels}

In general, there are two types of market participators: \emph{individuals} and \emph{institutions}. It is reasonable to believe that these two groups express their ``sentiment'' in different ways. As we could imagine, the social media are more individual-oriented as it is more casual and the institutional investors seldom express their attitude towards market directly in public. As remarked by \citet{VermaSoydemir}, even a survey for institutions may contain biases since they could deviate heavily from what they published. It is obvious, however, that sophisticated investors like institutions constitute the majority in contributing to the derivatives market. As \citet{EasleyOHaraSrinivas} point out, the informed traders are more likely to trade in the option market rather than in the equity market. Therefore, we tend to interpret the sentiment extracted from the social media as \emph{individual} investors' sentiment, while treat the option-implied one as \emph{institutional} investors' attitude towards the market. We expect these two to be significantly different. Finally, since the overall market is made up of and also traded by both individuals and institutions, the market-type proxy could be interpreted as the sentiment for the whole market.

In order to have a more comprehensive understanding about the financial sentiment, it is natural to consider all three channels simultaneously. The equal-weighted sum as an overall sentiment index is the simplest but may cause information loss, since the three indices are not fully inter-independent, as shown by the correlation calculation given later in Table \ref{Correlation}. Another possible \emph{linear} combination of the three indices could be figured out by VAR when addressing the predictability issue. However, the most interesting mixture could be a \emph{nonlinear} form through a neural network as discussed further in the next section.

\section{Stock return predictability by sentiment indices}

Investigation on how to predict the future stock return requires better time-series analysis tools and it still remains challenging. \cite{VermaSoydemir} studied the predicting ability of investors' sentiment, with the presence of other fundamental market factors like Fama-French three factors \citep[see][]{fama93commonrisk}, through the Vector Autoregression (VAR) which is a basic model in econometrics.  VAR, though simple and clear enough, only captures the linear relationship among different time-series data. In this paper, we propose to use Long Short-Term Memory (LSTM) model to analyze the predictability of different sentiment indices on stock return, as it could capture the nonlinear features that traditional VAR fails to include. Our testing results conclude that the LSTM model outperforms VAR in a yearly basis in terms of lower mean square error.

\subsection{Basic statistics}


Table \ref{Correlation} summarizes, for each individual stock, the correlation coefficients between any two quantities out of three different sentiment indices themselves. We also do the similar analysis for sentiment index at time $t$ with stock return at $t+1$, $r_{t+1}^i$.

\begin{table*}[!htbp]
\caption{Correlation coefficient between two quantities for each individual stock, where $BSI^i$ stands for our BERT-based sentiment index for stock $i$, $OSI^i$ for option-implied sentiment index, and $MSI^i$ for market-implied one; $r_{t+1}^i$ represents stock return at $t+1$.}
\centering
\begin{tabular}{ccccccc}
\toprule
Stock $i$    &&& Tencent (0700.HK) && CCB (0939.HK) & Ping An (2318.HK)\\
\midrule
\multicolumn{7}{c}{\emph{Correlation between different sentiment indices for each stock}}\\
\midrule
$BSI^i$ v.s. $OSI^i$ &&& 0.0347  && -0.0026 & -0.0448 \\
$BSI^i$ v.s. $MSI^i$ &&& -0.3442 && 0.1944  & 0.2024  \\
$OSI^i$ v.s. $MSI^i$ &&& -0.1776 && 0.1463  & 0.0116  \\
\midrule
\multicolumn{7}{c}{\emph{Correlation between today's sentiment index and tomorrow's stock return}}\\
\midrule
$BSI^i_t$ v.s. $r_{t+1}^i$   &&& -0.0205 && -0.0387 & 0.0710  \\
$OSI^i_t$ v.s. $r_{t+1}^i$   &&& -0.0052 && -0.0094 & -0.0327 \\
$MSI^i_t$ v.s. $r_{t+1}^i$   &&& -0.0304 && -0.0068 & 0.0337  \\
\bottomrule
\end{tabular}
\label{Correlation}
\end{table*}

From this table we can see that there does not exist a persistent linear relationship across different quantities in the individual stock level. For instance, when we compare $BSI^i$ with $MSI^i$, it could have positive or negative correlations across different stocks, though relatively strong in magnitude. Given a certain individual, the relation between different pairs of sentiment indices looks borderline as well. As the simple prediction power check, all values of correlation coefficients seem low, which may indicate a hidden nonlinear affection of sentiment on future stock return.

\subsection{Predictability of sentiment indices on stock return}

In this subsection, we examine whether our BERT-based financial sentiment index and the other two indices could predict the market or not, especially on predicting the future stock return. This is done by further considering the presence of the classical risk factors which have been proved to have pricing power on stocks. Following the work by \citet{VermaSoydemir}, we select eight fundamental factors as \emph{control variables}, including one-month interest rate ($r_1$), economic risk premium defined by the difference between three-month and one-month interest rates ($r_3-r_1$), inflation rate ($Inf$), the return on portfolio of winning stocks over past twelve months minus those losing stocks ($UMD$), the currency fluctuation of Hong Kong dollar ($HKD$), and the Fama-French three factors: the excess market portfolio return ($r_m-r_1$), the return on portfolio of small companies minus big ones ($SMB$), and the return on portfolio of high book to market value companies minus low book to market value ones ($HML$). The full time period in our experiment covers from January 1, 2016 to December 31, 2018. The individual stock return is defined by its \emph{log return}, namely, $r_t^i = \log(S_t^i/S_{t-1}^i)$ where $S_t^i$ is the price for stock $i$ at time $t$. All time-series data are normalized and scaled to have mean 0 and variance 1.
%

\subsubsection{VAR and LSTM modelling}

We first employ VAR as a traditional time-series analysis tool to investigate the predictability of the sentiment on future stock return. More precisely, we consider the following model,
\begin{equation}
	Y_t^i = A^i + \sum^\ell_{s=1} B_s^i Y_{t-s}^i + \boldsymbol\epsilon_t^i
\label{VAR}
\end{equation}
where $Y_t^i$ is a column vector for stock $i$ consisting of variables which we believe could have an inter-temporal relationship (in our case it contains stock return and different sentiment indices with risky factors above), $A^i$ is a time-invariant constant term, $\ell$ is the look-backward length (which is set to be 2 here), taking into account the possible time lag effect of sentiment, $B_s^i$ is the matrix of coefficients for the $s$-lag vector $Y_{t-s}^i$, and $\boldsymbol\epsilon_t^i$ is the error term. Note that (\ref{VAR}) can be written in an ordinary \emph{linear regression} form
\begin{equation}
y^i_{tm} = a_m^i + \sum^\ell_{s=1} \sum^{N^i}_{n=1} b^i_{sn} y^i_{(t-s)n} + \epsilon^i_{tm},
\label{VAR_linear}
\end{equation}
where $y^i_{tm}$ is our concerned component of $Y_t^i$ (that is stock return which we are going to predict) for stock $i$, and $y^i_{(t-s)n}$ is the $n$th-component of $N^i$-dimensional vector $Y_{t-s}^i$ with corresponding coefficient $b^i_{sn}$.

Despite VAR always acts as a benchmark forecasting model in finance, it requires strong model assumptions, like Gaussian white noise and dependence of predetermined variables. We emphasize that VAR is a linear prediction model as evidenced from (\ref{VAR_linear}). We now adopt LSTM as a powerful machine learning method in predicting future based on past information without assuming any noise form. Most importantly, LSTM could capture possible \emph{nonlinear} features behind the time series. The hyperparameters of our LSTM model include the number of layers $L$ and the number of training epochs $E$. Moreover, we keep the same maximum time lag $\ell$ and set the hidden size to be the same number of independent variables of VAR above, in order to have a proper comparison. Note that, since a linear structure is actually a special case of a feedforward neural network when armed with a linear transformer, one should expect that LSTM performs at least as good as VAR.

\subsubsection{Predictability testing results}

In our experiments of each individual stock return prediction, \emph{dates} within a calendar year are randomly distributed into the \emph{training} set $D_{tr}^i$ and the \emph{test} set $D_{te}^i$, with proportion 80\% and 20\%, respectively, which are shared for both VAR and LSTM models fitting in parallel. Namely, the stock return $r_t^i$ such that $t\in D_{tr}^i$ as output together with sentiment indices and all other factors at time $t-2$ and $t-1$ as inputs are used to train our models, and we choose to report mean square error (MSE) not only on $D_{te}^i$ but also on $D_{tr}^i$ as well as the whole year set $D_{wh}^i$. Note that we let three sentiment indices enter into inputs separately and also all together for different experiments in order to see whether the combination of different sentimental sources could enhance the prediction further or not. Besides, the reason why we start with the yearly basis testing is that the influence of sentiment, though may maintain for a while, cannot last for too long.

Table \ref{2016_2018} lists the yearly-basis prediction accuracy of different sentiment indices and their mixture (with the presence of other factors mentioned above), in terms of MSE between real stock returns and predicted ones calculated on different sets of dates under both LSTM and VAR (in bracket) models, for three selected stocks, respectively. We can see that all MSE's calculated from LSTM are smaller than those from VAR in the table (including on test sets $D_{te}^i$'s), indicating that the yearly-basis stock return prediction from LSTM performs in general better than using traditional time-series tool VAR. In other words, investors' sentiment predicts the market more likely in a \emph{nonlinear} fashion.

\begin{table*}[!htbp]
\captionsetup{font={small}}
\caption{Individual stock return prediction (of different sentiment indices and their mixture with presence of other factors) accuracy in terms of MSE (unit $\times10^{-5}$) based on LSTM and VAR (in bracket) for 2016, 2017, and 2018, respectively; \textbf{bold} numbers show the best sentiment index as a predictor on a certain set of dates in a particular year; the case without using any sentiment index to predict the market is also examined.}
\centering
\resizebox{1\textwidth}{!}{
\begin{tabular}{cccccccccc}
\toprule
Stock $i$ & \multicolumn{3}{c}{Tencent (0700.HK)} & \multicolumn{3}{c}{CCB (0939.HK)} & \multicolumn{3}{c}{Ping An (2318.HK)}\\
\midrule
MSE on   & $D_{tr}^{0700}$ & $D_{te}^{0700}$ & $D_{wh}^{0700}$ & $D_{tr}^{0939}$ & $D_{te}^{0939}$ & $D_{wh}^{0939}$ & $D_{tr}^{2318}$ & $D_{te}^{2318}$ & $D_{wh}^{2318}$\\
\midrule
\midrule
\multicolumn{10}{l}{\emph{For 2016}}\\
\midrule
$BSI^i$      & 2.87 (4.04) & 2.91 (4.64) & 2.88 (4.16) & 2.51 (3.55) & 3.06 (\textbf{5.01}) & 2.62 (\textbf{3.84}) & 2.98 (4.47) & 4.17 (\textbf{6.66}) & 3.22 (4.90)\\
$OSI^i$      & 2.83 (3.99) & 2.97 (4.67) & 2.85 (4.13) & 2.56 (3.57) & 3.06 (5.09) & 2.66 (3.87) & 3.76 (4.64) & 4.44 (6.71) & 3.90 (5.05)\\
$MSI^i$      & 2.58 (4.00) & 2.52 (\textbf{4.62}) & 2.57 (4.12) & 2.56 (3.56) & 3.05 (5.06) & 2.66 (3.86) & 3.75 (4.62) & 4.59 (6.69) & 3.91 (5.03)\\
\emph{Mixture}      & \textbf{2.40} (\textbf{3.90}) & \textbf{2.34} (4.83) & \textbf{2.39} (\textbf{4.08}) & \textbf{2.17} (\textbf{3.52}) & \textbf{2.81} (5.26) & \textbf{2.29} (3.86) & \textbf{2.52} (\textbf{4.41}) & \textbf{3.28} (6.78) & \textbf{2.67} (\textbf{4.88})\\
\midrule
\emph{No-SI}		 & 3.88 (3.86) & 5.56 (5.53)   & 4.21 (4.19) & 3.62 (3.60) & 4.96 (5.06) & 3.89 (3.89) & 4.77 (4.58) & 6.80 (7.26) & 5.17 (5.11) \\
\midrule
\midrule
\multicolumn{10}{l}{\emph{For 2017}}\\
\midrule
$BSI^i$      & 2.12 (3.41) & 2.24 (5.32) & 2.14 (3.79) & 1.56 (2.58) & 1.52 (3.27) & 1.55 (2.72) & 2.57 (4.35) & \textbf{2.01} (\textbf{4.97}) & 2.46 (4.48) \\
$OSI^i$      & 2.07 (3.42) & \textbf{2.16} (5.33) & \textbf{2.09} (3.80) & 1.65 (2.60) & 1.52 (3.32) & 1.63 (2.74) & 2.80 (4.47) & 2.09 (4.97) & 2.66 (4.57) \\
$MSI^i$      & 2.32 (3.38) & 2.33 (\textbf{5.30}) & 2.32 (3.76) & 1.75 (2.57) & 1.52 (\textbf{3.30}) & 1.70 (\textbf{2.71}) & 2.97 (4.45) & 2.52 (5.04) & 2.88 (4.57) \\
\emph{Mixture}      & \textbf{2.05} (\textbf{3.32}) & 2.29 (5.34) & 2.10 (\textbf{3.72}) & \textbf{1.45} (\textbf{2.54}) & \textbf{1.46} (3.41)          & \textbf{1.45} (2.72) & \textbf{2.48} (\textbf{4.30}) & 2.07 (5.07) & \textbf{2.40} (\textbf{4.46}) \\
\midrule
\emph{No-SI}		 & 2.97 (3.60) & 4.45 (4.43)   & 3.27 (3.77) & 2.29 (2.68) & 2.96 (2.98) & 2.42 (2.74) & 3.90 (4.51) & 5.30 (4.84) & 4.18 (4.58) \\
\midrule
\midrule
\multicolumn{10}{l}{\emph{For 2018}}\\
\midrule
$BSI^i$      & 5.79 (8.89) & 5.52 (11.88) & 5.74 (9.49) & 2.99 (4.54) & 2.83 (5.10) & 2.96 (4.65) & 2.59 (4.51) & 2.23 (5.51) & 2.52 (4.71) \\
$OSI^i$      & 6.14 (8.84) & 4.85 (\textbf{11.33}) & 5.88 (9.34) & 3.34 (4.49) & 2.95 (\textbf{5.04}) & 3.27 (4.60) & 2.91 (4.48) & 2.75 (5.58) & 2.88 (4.70)\\
$MSI^i$      & 6.20 (8.95) & 5.67 (11.67) & 6.09 (9.49) & 2.85 (4.50) & 2.71 (5.06) & 2.83 (4.61) & 2.95 (4.42) & 2.54 (\textbf{5.43}) & 2.86 (4.62) \\
\emph{Mixture}      & \textbf{5.33} (\textbf{8.68}) & \textbf{4.40} (11.91) & \textbf{5.14} (\textbf{9.32}) & \textbf{2.58} (\textbf{4.43}) & \textbf{2.31} (5.13) & \textbf{2.53} (\textbf{4.57}) & \textbf{2.27} (\textbf{4.33}) & \textbf{2.04} (5.57) & \textbf{2.22} (\textbf{4.58})\\
\midrule
\emph{No-SI}		 & 7.20 (9.01) & 11.93 (12.17) & 8.14 (9.64) & 4.03 (4.42) & 5.93 (5.78) & 4.41 (4.69) & 3.79 (4.51) & 5.45 (5.81) & 4.12 (4.77) \\
\bottomrule
\end{tabular}
}
\label{2016_2018}
\end{table*}

Another worth-mentioning observation is that the more complex combination of three sentiment indices leads to a better prediction, as almost all the mixtures have lower MSE's than the case when there is only one sentiment index added under LSTM setting (as we can find from the table that bold numbers, which stands for the \emph{best} sentiment predictor within a certain set of dates, on LSTM positions always appear in the \emph{Mixture} lines), while it is not always true for VAR model. This result confirms some complicated influence structure of sentiment on stock return in the individual level and suggests that the combination of different channels does help to improve the accuracy of the predicting power of financial sentiment. As a supplement, we also examine the simpler case that leverages \emph{only} eight factors to predict the market without utilizing any sentiment index (denoted by \emph{No-SI}). As we could imagine, adding sentimental factors is indeed valuable under LSTM model, since the appearance of sentiment indices as predictors significantly reduces the predicting error, reflecting on lower MSE's in the Table \ref{2016_2018}.

\begin{figure*}[!htbp]
\centering
\subfigure
{
\includegraphics[width=4.4cm]{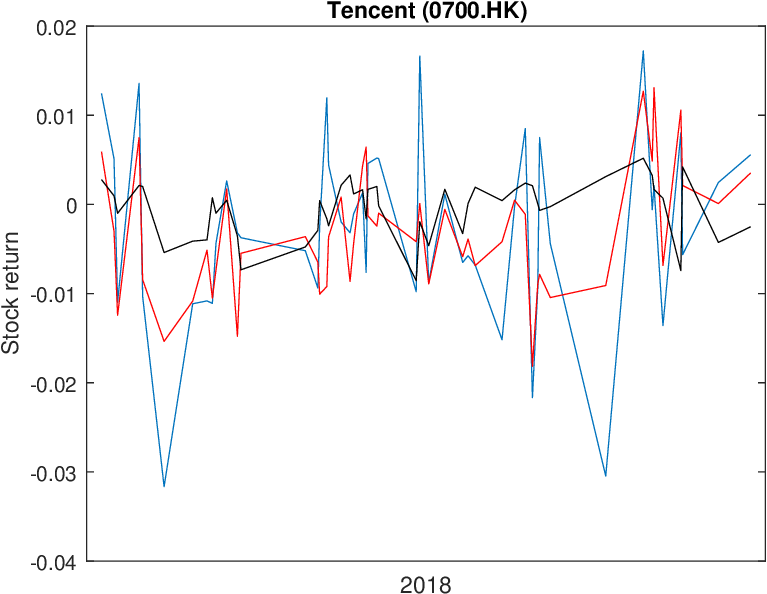}
\label{Tencent_LSTM_vs_VAR}
}
\subfigure
{
\includegraphics[width=4.4cm]{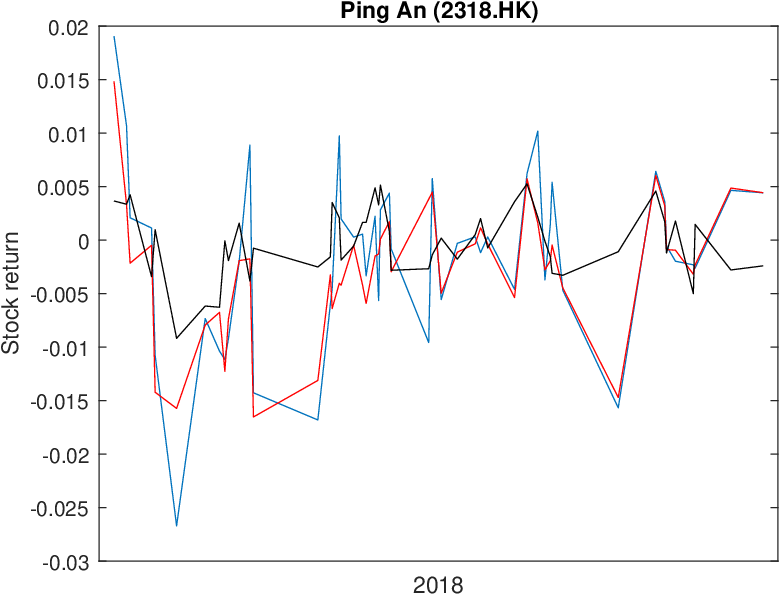}
\label{PingAn_LSTM_vs_VAR}
}
\subfigure
{
\includegraphics[width=4.4cm]{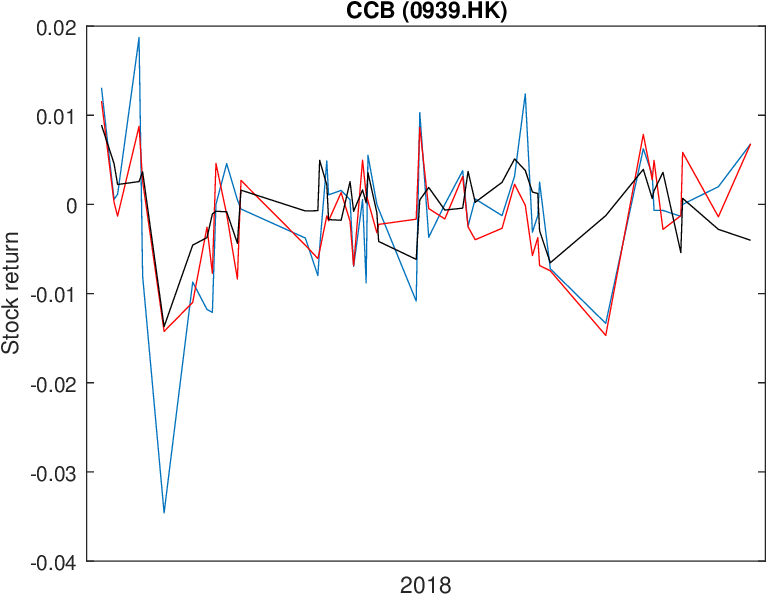}
\label{CCB_LSTM_vs_VAR}
}

\subfigure
{
\includegraphics[width=4.4cm]{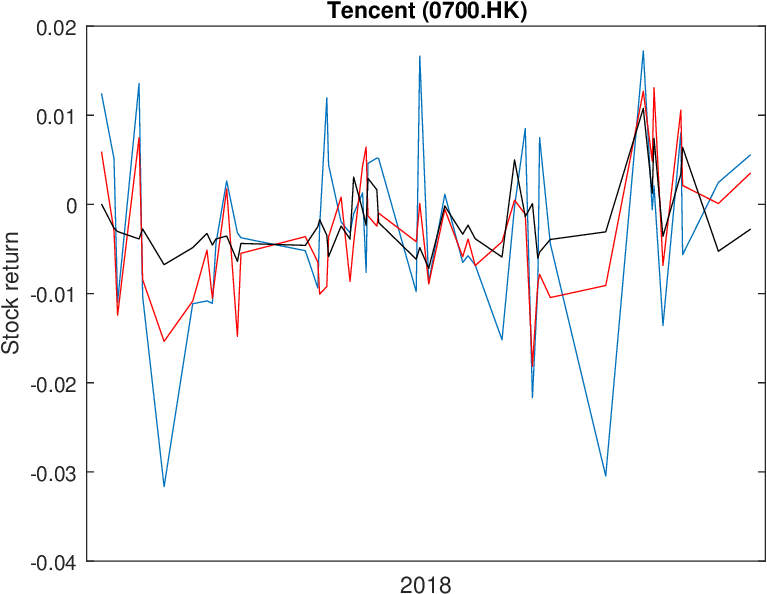}
\label{Tencent_LSTM_only}
}
\subfigure
{
\includegraphics[width=4.4cm]{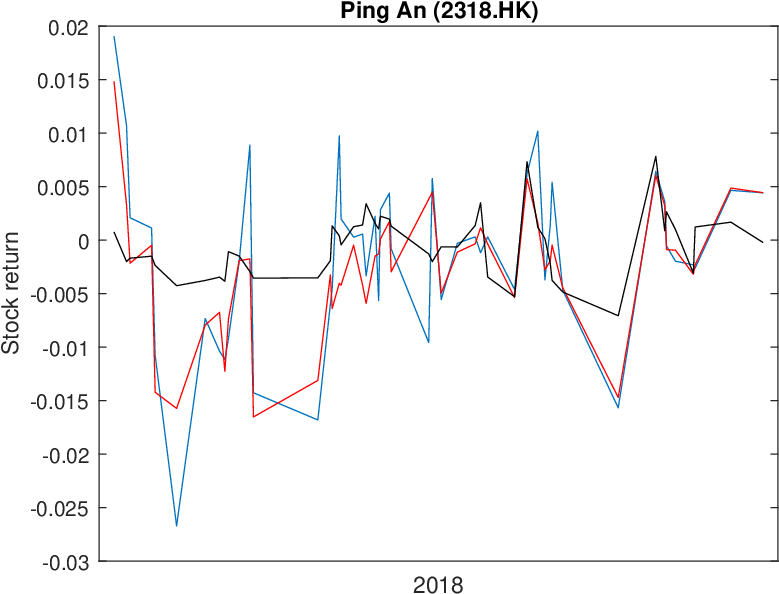}
\label{PingAn_LSTM_only}
}
\subfigure
{
\includegraphics[width=4.4cm]{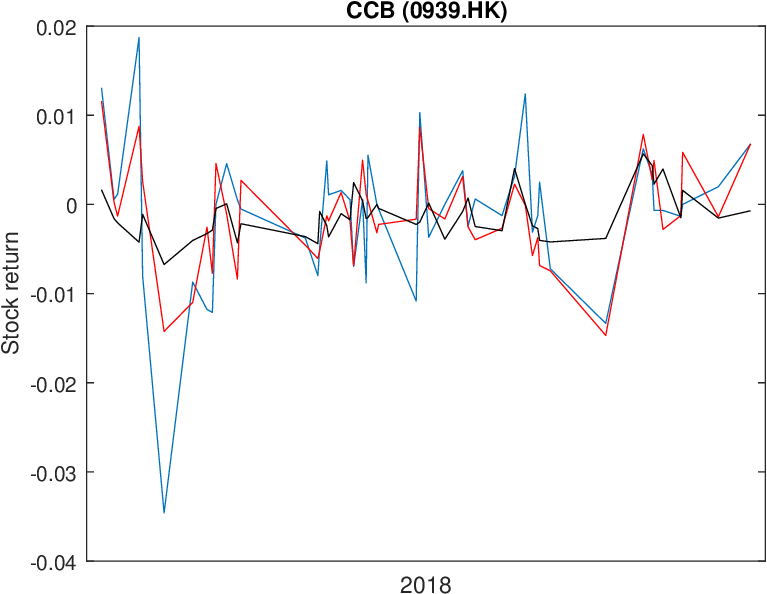}
\label{CCB_LSTM_only}
}
\captionsetup{font={small}}
\caption{Real stock return movement (in blue) versus predicted ones from different models on test set of 2018 for each individual stock. The first row displays prediction of \emph{Mixture} (with presence of other factors) from LSTM (in red) and VAR (in black), while the second row exhibits prediction of \emph{Mixture} (in red) and \emph{No-SI} (in black) for LSTM model only; among all figures \emph{Mixture} under LSTM setting performs best on prediction of stock return in the sense of closer distance to the real one.}
\label{LSTM_VAR}
\end{figure*}

The above conclusion can also be found graphically in Figure \ref{LSTM_VAR}, where as an example we take \emph{Mixture} as the sentimental predictor along the dates of test set in 2018 (while the other sentiment indices under the rest of years behave similarly in our experiments). The first row of Figure \ref{LSTM_VAR} shows, stock by stock, the real stock return movement (blue line) versus \emph{Mixture}-predicted returns (under the attendance of those considered risk factors). The prediction by LSTM is depicted in red while by VAR in black. It is apparent that LSTM prediction is much closer to the real return than VAR prediction for any individual stock. We also display the comparison between whether we utilize sentiment as predictor or not in LSTM model, as exhibited by the second row in the figure. We can still see the dominating performance of \emph{Mixture} (in red) over \emph{No-SI} (in black) on predicting the real return fluctuation (in blue) for each stock. Note that since we randomly assign dates into test set, the curves in figures are accomplished through connecting real or predicted return data points on dates of test set in a correct time order.

As for the predictability testing for the whole time period from 2016 to 2018, we find that LSTM performs worse than VAR on the test sets for all three stocks, even with the same hyperparameters used for yearly-based testing, as shown by the underlined numbers in Table \ref{whole_period}. This is possibly because the ML model is over-fitted. To see this, let us design another trivial model that always predicts zero return for any stock, and then we still calculate the MSE between real stock return and constant zero, the result of which is summarized in Table \ref{constant_zero}. Since the prediction accuracy is not far away from those of LSTM and VAR, it seems that any attempt to predict the future stock return over a \emph{longer} time period is in vain, as if the model is learning a random signal with zero mean.

\begin{table*}[!htbp]
\captionsetup{font={small}}
\caption{Individual stock return prediction (of different sentiment indices and their mixture with the presence of other factors) accuracy in terms of MSE (unit: $\times10^{-5}$) based on LSTM and VAR (in bracket) models for the complete time period from 2016 to 2018; \underline{underline} numbers show where VAR outperforms LSTM.}
\centering
\resizebox{1\textwidth}{!}{
\begin{tabular}{cccccccccc}
\toprule
Stock $i$ & \multicolumn{3}{c}{Tencent (0700.HK)} & \multicolumn{3}{c}{CCB (0939.HK)} & \multicolumn{3}{c}{Ping An (2318.HK)}\\
\midrule
MSE on   & training set & test set & whole set & training set & test set & whole set & training set & test set & whole set\\
\midrule
\midrule
\multicolumn{10}{l}{\emph{From 2016 to 2018}}\\
\midrule
$BSI^i$      & 5.05 (5.96) & 7.23 (\underline{6.54}) & 5.49 (6.07) & 3.32 (3.97) & 4.83 (\underline{4.49}) & 3.62 (4.08) & 4.45 (5.18) & 5.64 (\underline{5.23}) & 4.69 (5.19) \\
$OSI^i$      & 4.71 (5.95) & 7.61 (\underline{6.53}) & 5.29 (6.07) & 3.54 (3.97) & 4.70 (\underline{4.46}) & 3.78 (4.06) & 4.67 (5.16) & 5.56 (\underline{5.21}) & 4.85 (5.17) \\
$MSI^i$      & 5.10 (5.95) & 7.42 (\underline{6.59}) & 5.57 (6.08) & 3.43 (3.97) & 4.70 (\underline{4.48}) & 3.68 (4.08) & 4.69 (5.16) & 5.74 (\underline{5.22}) & 4.90 (5.18) \\
Mixture      & 4.52 (5.92) & 8.00 (\underline{6.61}) & 5.21 (6.06) & 3.29 (3.96) & 5.09 (\underline{4.52}) & 3.65 (4.07) & 4.36 (5.13) & 5.93 (\underline{5.23}) & 4.67 (5.15) \\
\bottomrule
\end{tabular}
}
\label{whole_period}
\end{table*}

\begin{table*}[!htbp]
\captionsetup{font={small}}
\caption{Individual stock return prediction accuracy in terms of MSE (unit: $\times10^{-5}$) based on a trivial model that always predicts constant zero stock return, for the complete time period from 2016 to 2018.}
\centering
\begin{tabular}{ccccc}
\toprule
Stock $i$    && Tencent (0700.HK) & CCB (0939.HK) & Ping An (2318.HK)\\
\midrule
Training set && 6.42  & 4.15 & 5.52 \\
Test set && 6.35 & 4.35  & 5.16  \\
Whole set && 6.40 & 4.19  & 5.45  \\
\bottomrule
\end{tabular}
\label{constant_zero}
\end{table*}

\section{Summary}

In this paper, we construct a textual financial sentiment index for three stocks that are listed in the Hong Kong Stock Exchange and have hot discussion on Weibo.com, using the state-of-the-art NLP model BERT developed by Google in recent years, as the first BERT-based sentiment index in the literature, to the best of our knowledge. We also demonstrate the dominating feature of our financial sentiment classification result over other existing deep learning methods in terms of precision, recall, and averaged F1 score. Apart from textual channel to extract investors' sentiment, the traditional approaches utilize the option data from derivatives markets and stock market data directly, resulting in option-implied and market-implied sentiment indices, respectively. Based on these three different information channels, we propose a more comprehensive framework for financial sentiment analysis by interpreting the textual sentiment as individual investors' emotion, and the option-implied one as institutional investors' opinion, while the market-implied one as overall attitude of all market participants. We also discuss the predictability of sentiment on stock return in the individual level. Rather than using traditional econometric methods like VAR, we adopt LSTM as an ML tool in order to capture the possible nonlinearity of sentiment impact on stock return. It turns out that LSTM performs better than VAR on prediction for a yearly basis in the sense of lower MSE's. However, when it comes to a longer time period, ML models seem easily over-fitted. How to deal with this issue could be our future research direction, and so is the extension of our BERT-based sentiment construction to a market-level analysis rather than individual level.

\bibliographystyle{ACM-Reference-Format}
\bibliography{BERT}

\appendix

\section{Descriptions on selected market data for market-implied sentiment index}
The appendix illustrates the five types of market data that are used to construct the market-implied sentiment index. Since these factors are all defined in a individual stock level, we ignore the superscript of stock $i$ here for simplicity.

\subsection{Relative strength index (RSI)}
RSI is usually interpreted as a signal of over purchasing or over selling. The value of it lies between 0 to 100 for which under 20 means over-sold and over 80 represents over-bought. It is calculated using the price changes over the past 14 days, which is given by
	\begin{equation}
		\text{RSI}_t = \frac{\sum^{14}_{i=1} (S_{t - i + 1} - S_{t - i})_+}{\sum^{14}_{i=1} (S_{t - i + 1} - S_{t - i})} \times 100
	\end{equation}
where $S_t$ stands for the stock at time $t$ and $(x)_+:= \max(x,0)$. Invented by Wilder 40 years ago, RSI is still an useful trading signal for investors nowadays, as indicated by \citet{MarekSediva}.

\subsection{Turnover ratio}

The turnover ratio is defined as the ratio of average trading volume (turnover) of past 10 days over that of past 250 days (almost the number of trading days in a year) normalized by a factor of 100, that is, the turnover ratio at $t$ is defined by
\begin{equation}
	\text{TR}_t = \frac{\bar{V}^{10}_t}{\bar{V}^{250}_t} \times 100
\end{equation}
where $\bar{V}^{10}_t$ denotes the average turnover of past 10 days standing at $t$, and similar for $\bar{V}^{250}_t$. As argued by \cite{ChongCaoWong}, turnover ratio is related to the willingness of investors involving in trading activities, with high value indicating a bull market while low for a bear market.

\subsection{Short-selling turnover ratio (SSTR)}

The SSTR is the ratio of short-selling volume (SSV) to the turnover ratio at time $t$, i.e.,
\begin{equation}
	\text{SSTR}_t = \frac{\text{SSV}_t}{\text{TR}_t}
\end{equation}
As short selling usually indicates a downside movement in the near future, we use this ratio as a proxy for negative information which could affect investors sentiment.

\subsection{Money flow}

Money flow is a more complicated indicator that captures the flow of the money when the stock is rising or falling. \citet{ChongCaoWong} mention that this value may reflect the trend of the stock as it reveals how the money is flowing in different market stages (rising or falling). Before defining the market flow, we first define the daily price (DP) as the average of highest (H), lowest (L) and closing prices (C) of a stock on day $t$:
\begin{equation}
	\text{DP}_t = \frac{1}{3}(\text{H}_t + \text{L}_t + \text{C}_t)
\end{equation}
Then we define the flow of money (FM) as:
\begin{equation}
	\text{FM}_t = \text{DP}_t \times \text{TR}_t.
\end{equation}
We say the flow of money is positive (PFM) if today's DP is higher than yesterday's, and similar for negative (NFM). Then the money flow (MF) is defined as the ratio of the accumulation of positive flow of money of past 30 days to the total value of both positive and negative ones, multiplied by 100
\begin{equation}
	\text{MF}_t = \frac{\text{PFM}^{30}_t}{\text{PFM}^{30}_t+\text{NFM}^{30}_t} \times 100
\end{equation}

\subsection{Put-call ratio}

The Put-Call (PC) ratio for an individual stock is defined by its number of put options (\#P) traded in the market during a given period over that of calls (\#C), i.e.,
\begin{equation}
	\text{PC}_t = \frac{\text{\#P}_t}{\text{\#C}_t}.
\end{equation}
As put option may indicate fall of stock price in the future whereas calls vice versa, the ratio can be viewed as the market direction in investors' expectation accordingly.

The result of principal component analysis on the above five factors within the considered time period of the main paper is our market-implied sentiment index.

\end{document}